\documentclass[11pt]{article}
\pdfoutput=1
\usepackage{jheppub}
\usepackage{amssymb,amsfonts}
\usepackage{mathrsfs}
\usepackage{amsmath}
\usepackage[nameinlink,capitalise,noabbrev]{cleveref}
\usepackage{tikz}
\usetikzlibrary{arrows.meta,positioning,calc}
\crefformat{section}{\S#2#1#3} 
\crefformat{subsection}{\S#2#1#3}
\crefformat{subsubsection}{\S#2#1#3}
\let\savenumberline\numberline
\def\numberline#1{\savenumberline{#1.}}
\makeatletter
\renewcommand{\@seccntformat}[1]{\csname the#1\endcsname.\,\,}
\makeatother
\renewcommand{\hat}[1]{\widehat{#1}}
\newcommand{\be}{\begin{equation}}
\newcommand{\ee}{\end{equation}}
\newcommand{\bea}{\begin{eqnarray}}
\newcommand{\eea}{\end{eqnarray}}

\makeatletter
\def\@fpheader{\relax}
\makeatother
\usepackage{graphicx}
\usepackage{latexsym}
\usepackage{orcidlink}

\title{Tropical Open Strings in a Kalb--Ramond Background}
\author[a,\orcidlink{0000-0002-4535-3198}]{Sarthak Duary,}
\emailAdd{sarthakduary@tsinghua.edu.cn}

\author[b,\orcidlink{0009-0005-0546-7894}]{Vi Hong,}
\emailAdd{vihong14@berkeley.edu}

\author[c,\orcidlink{0009-0008-2963-2497}]{Sourav Maji}
\emailAdd{souravmaji@hri.res.in}

\affiliation[a]{Yau Mathematical Sciences Center (YMSC), Tsinghua University, \\ Haidian District, Beijing 100084, China}

\affiliation[b]{Leinweber Institute for Theoretical Physics and Department of Physics,
\\University of California, Berkeley, CA 94720-7300, USA}

\affiliation[c]{Harish-Chandra Research Institute, A CI of Homi Bhabha National Institute,
\\Chhatnag Road, Jhunsi, Prayagraj (Allahabad), Uttar Pradesh 211019, India}
\abstract{\begin{abstract}

We study tropical open strings arising from the analytically continued action of tropological sigma models in the presence of a constant Kalb--Ramond background. We show that the background field separates the theory into a generic sector and a critical sector. In the generic sector, the canonical structure produces a kinematical nonlocal bracket between the transverse and leafwise target-space coordinates. At the critical value of the Kalb--Ramond field, the Legendre map degenerates and the theory develops a primary constraint, requiring quantization by the Dirac--Bergmann procedure. This critical sector has a finite reduced phase-space structure and differs from the naive limit of the generic theory.
\end{abstract}
}
\begin{document}
\maketitle

\section{Introduction}
\label{sec:Intro}

String theory has been remarkably successful in describing perturbative scattering processes and equilibrium backgrounds. However, a fully developed worldsheet formulation of string theory out of equilibrium has yet to be constructed. It is not clear what geometric structures should replace the usual relativistic worldsheet or whether or not such a formalism even exists in the first place. In nonequilibrium quantum field theory, real-time evolution is naturally formulated using the Schwinger--Keldysh, or closed-time, contour. In \cite{neq, ssk, keq}, using the techniques of large-$N$ dualities, the authors studied how the closed-time contour in non-equilibrium field theory should be generalized to perturbative string theory. The predicted worldsheet, on the Schwinger Keldysh time contour, is expected to decompose into a triple,
\begin{align}
    \Sigma = \Sigma^{+} \cup \Sigma^{\wedge} \cup \Sigma^{-},
\end{align}
with $\Sigma^{+}$and $\Sigma^{-}$ corresponding to the forward and backward branches, respectively, and the ``wedge region” $\Sigma^{\wedge}$ connecting them at the turnaround. 

The wedge region exhibits strange properties that prevent us from immediately constructing the worldsheet path integral. It is expected to be highly anisotropic described by non-relativistic foliations whose leaves connect boundary data from the forward and the backward branches. Though the wedge region is expected to admit its own genus expansion, it is not known what exotic topologies make up this sum. As a result, we expect that the geometry of the wedge region requires tools beyond the usual relativistic worldsheet description.

In \cite{trsm}, progress was made to describe the anisotropic behavior of the wedge region by utilizing the tools of tropical geometry. There, a new class of cohomological field theories, called tropological sigma models, was constructed by taking a tropical limit of the topological A-model. In these theories, the complex structure of the worldsheet degenerates into a nilpotent endomorphism of the tangent bundle $T\Sigma$. The endomorphism defines a rank-one distribution and thus equips the worldsheet with a foliation.

In \cite{Albrychiewicz:2024tqe}, it was found that solutions to the tropological sigma models naturally gave rise to a non-relativistic brane. An explicit Hamiltonian description was constructed and thus, provided a framework to construct states and observables for the resultant non-relativistic theory. These results suggest that tropological sigma models are connected to a broader class of non-relativistic theories, including Carrollian field theories and approaches to flat-space holography \cite{Bergshoeff:2020xhv, Kluson:2022jxh, Bergshoeff:2023rkk}. More recently, related Carrollian boundary symmetry algebras have been studied from a representation-theoretic perspective in which one of the related boundary-preserving subalgebras was identified with the worldsheet symmetry algebra of tropological open strings \cite{Buzaglo:2025bcca}. 

The foliated and degenerate structure of the wedge region and the subsequent models inspired by it, suggests that a more complete description may involve nonlocal or noncommtative structures. A standard setting where noncommutative geometry appears in string theory is that of ordinary open-string theory propagating in a constant Kalb--Ramond background \cite{Abouelsaood:1986gd}. Subsequent works showed that the endpoint coordinates of these open strings become noncommutative \cite{Ardalan:1998ce, ChuHo:1998wk, Schomerus:1999ug}. In the usual Seiberg--Witten analysis, a background $B$-field induces a noncommutative algebra of endpoint coordinates and leads to a noncommutative gauge theory on the D-brane worldvolume \cite{Seiberg:1999vs}. This was remarkable as it provided an explicit example of an effective noncommutative spacetime geometry. In the intrinsically tensionless string theory, the canonical kinetic contribution collapses and the constant Kalb--Ramond coupling becomes the unique symplectic datum on the reduced boundary phase space; inverting the resulting boundary symplectic form yields the noncommutative endpoint algebra, whose noncommutativity parameter matches the tensionless limit of the  Seiberg--Witten analysis \cite{Duary:2025hdb}.

In the present work, we investigate the effects of a constant $B$-field on a tropical open string. In \Cref{sec:tropological-openstringreview}, we review the relevant features of tropological open strings and their analytically continued action, and the boundary conditions that define the corresponding nonrelativistic tropical branes \cite{trsm,Albrychiewicz:2024tqe}.

In \Cref{sec:Kalb_ramond_bacgrd}, we place the tropical open string in a constant Kalb--Ramond field. We find that the tropical case exhibits several interesting properties. First, there are two distinct sectors of the theory. One for generic values of constant $B$-field as well as a separate sector for a single critical value of the Kalb--Ramond field. Interestingly, at this critical value, the endpoints decouple from the brane and the Legendre transform drops rank. The theory has an extra primary constraint and thus must be treated carefully with the Dirac--Bergmann procedure for constrained Hamiltonian systems \cite{Dirac:1950pj,Bergmann:1949zz,htbook}. In both sectors, the commutator exhibits nonlocal behavior but this nonlocality is not dynamical, but rather a kinematical consequence of the first-order symplectic structure. The behavior differs in comparison with the Seiberg--Witten mechanism. The analytically continued tropical action already contains an intrinsic first-order coupling between the transverse coordinate and the leaf coordinate. When a constant Kalb--Ramond field is introduced, this coupling is rescaled rather than being reintroduced as in the standard Seiberg--Witten mechanism. As a result, the decoupling that occurs at the critical sector is not a large-$B$ limit but rather a cancellation of this term that generates a primary constraint. 

In \Cref{sec:can_sym_form}, we rederive the same endpoint noncommutativity from the canonical symplectic form and explain how the $B$-field rescales the boundary symplectic structure. In \Cref{sec:con}, we conclude with some open questions.

\section{Review of Tropological Open Strings}
\label{sec:tropological-openstringreview}
In this section, we review the construction of the tropological sigma models and the open-string solutions associated to the analytic continued theory. We restrict the discussion to the relevant details needed for the analysis of the Kalb--Ramond coupling in \Cref{sec:Kalb_ramond_bacgrd}. A more detailed review of tropological sigma models can be found here \cite{trsm}.

\subsection{Tropological Sigma Models}

We begin with the topological $A$-model obtained by twisting a standard relativistic two-dimensional sigma model with $\mathcal{N}=(2,2)$ supersymmetry \cite{ewtsm,Witten:1991zz}. Let $\Sigma$ denote a Riemannian worldsheet and let $M$ be the target space. We write the worldsheet complex structure as $\hat{\varepsilon}$, worldsheet coordinates as $\sigma^\alpha$, and the worldsheet metric as $\hat{g}$. Locally, the maps $\Phi$, from worldsheet to target space, are given by the coordinate representatives $Y^i(\sigma)$ and are written as
\begin{equation}
\Phi:\Sigma\to M,
\qquad
\Phi(\sigma)=Y^i(\sigma).
\end{equation}
We use a hat to distinguish these geometric structures from their tropicalized counterpart, consistent with the notation introduced in \cite{trsm}. For simplicity, we assume that $M$ is a smooth manifold of real dimension 2 equipped with an almost complex structure $\hat{J}$. 

By construction, the A-model localizes on pseudoholomorphic maps which are given in local coordinates as
\begin{align}
\label{eqn:LocEqn}
    \hat{E}_\alpha^{\;\;i}=\hat{\varepsilon}_{\alpha}^{\;\;\beta}\partial_\beta Y^i - \hat{J}_j^{\;\;i}\partial_\alpha Y^j=0.
\end{align}
We take  $(r,\theta)$ and $(X,\Theta)$ as adapted worldsheet and target space polar coordinates.  Then we may define the subtropical deformation on the worldsheet and target space \cite{viro} as, 
\begin{align}
\label{eqn:ViroSubTrop}
    z=\exp\left\{\frac{r}{\hbar}+i\theta\right\}, \quad Z=\exp\left\{\frac{X}{\hbar}+i\Theta\right\}.
\end{align}
The tropical limit is best understood as a Maslov (also called Litvinov-Maslov) dequantization \cite{msintro, litvinov, viro, virohyper} of the underlying coordinates and is achieved by taking  $\hbar\rightarrow 0$ of objects that depend on \eqref{eqn:ViroSubTrop}. In this limit, the complex structures defined on both the worldsheet and the target space now degenerate to rank-one nilpotent endomorphisms given by
\begin{align}
\label{eqn:TropJor}
    \varepsilon_{\alpha}^{\;\;\beta}d\sigma^\alpha \otimes \frac{\partial}{\partial \sigma^\beta}=dr \otimes\frac{\partial}{\partial\theta}, \quad J_{i}^{\;\;j}dY^i\otimes\frac{\partial}{\partial Y^j}=dX\otimes\frac{\partial}{\partial \Theta}.
\end{align}
The original condition that complex structures squares to minus the identity now is replaced by the degenerate nilpotency condition. These nilpotent structures define rank-one distributions on the worldsheet and on the target space. Since the distributions are one-dimensional and involutive, they thus foliate the worldsheet and the target space such that the leaves are generated by $\partial_\theta$ and $\partial_\Theta$. The worldsheet direction $r$ is transverse to the leaves on the worldsheet and likewise the target space coordinate $X$ is transverse to the leaves on the target space. 

The localization equations degenerate to 
\begin{align}
\label{eqn:LocEqnAdaptedCoord}
    E_r^{\;\;X}=\partial_\theta X=0, \quad E_r^{\;\;\Theta}=\partial_\theta\Theta-\partial_r X=0, \quad E_{\theta}^{\;\;X}=0, \quad E_{\theta}^{\;\;\Theta}=-\partial_\theta X=0. 
    \end{align}
General local solutions are then given as 
\begin{align}
    X(r,\theta)&=X_0(r), \\
    \Theta(r,\theta)&=\Theta_0(r)+\theta\partial_rX_0(r),
\end{align}
where $X_0(r)$ and $\Theta_0(r)$ are differentiable functions of $r$ only. Here, the transverse coordinate $X$ is constant along the worldsheet leaves while the leaf coordinate $\Theta$ varies according to the transverse derivative of $X$. This coupling between transverse and leafwise directions will play an important role after analytic continuation when we study the effects by the Kalb--Ramond coupling.

The tropical localization equations are additionally invariant under a new type of symmetry that is not present in the relativistic case. Following the nomenclature of \cite{trsm}, we call it an $\alpha$ symmetry. The infinitesimal variations of fields $X, \Theta$ that preserve localization equations \eqref{eqn:LocEqnAdaptedCoord} are
\begin{align}
\label{eqn:AlphaSym1}
    \delta X&=\alpha_1(r),\\
\label{eqn:AlphaSym2}    
    \delta \Theta&=\alpha_0(r)+\theta\partial_r\alpha_1(r),
\end{align}
where $\alpha_0(r)$ and $\alpha_1(r)$ are arbitrary functions of the transverse coordinate. 

In the quantization procedure, the $\alpha$ symmetry is interpreted as a gauge symmetry that one has to fix to impose the appropriate momenta constraints. It acts on the functions $X_0(r)$ and $\Theta_0(r)$ that parametrize the local solutions and must therefore be treated when defining the physical phase space. 

Following the BRST construction of the theory \cite{ewcoho,trsm} and after integrating out the term quadratic in $B$, the bosonic part of the action is
\begin{align}
\label{eqn:TropAction}
    S=\int_{\Sigma} dr d\theta\left\{\frac{1}{2}(\partial_\theta\Theta-\partial_r X)^2+\beta\partial_\theta X\right\},
\end{align}
where $\beta$ is a Lagrange-multiplier field.

\subsection{An Analytic Continuation}
We now review the analytically continued theory whose open-string solutions give rise to tropical branes \cite{Albrychiewicz:2024tqe}. Starting from the bosonic sector of the tropological sigma models \cite{trsm}, one obtains a theory with propagating local degrees of freedom by analytically continuing the leaf coordinate
\begin{align}
\label{eqn:Analytcont}
    \theta \rightarrow it.   
\end{align} 
Directly plugging in \eqref{eqn:Analytcont} into the bosonic action results in an expression that is not automatically a real Lorentzian action. In order to preserve unitarity by keeping the action real-valued, the action is modified via a topoogical invariant given by the pull-back of the two form $dX\wedge d\Theta$ to $\Sigma$ \cite{Albrychiewicz:2024tqe}. The resulting action is 
\begin{align}
S \;=\; \int dt\,dr \left[
\frac{1}{2}\big(\partial_t \Theta\big)^2
-\frac{1}{2}\big(\partial_r X\big)^2
+\big(\beta - \partial_r \Theta\big)\,\partial_t X
\right].
\end{align}
Varying the action with respect to $X$, $\Theta$, and $\beta$ gives the bulk equations of motion
\begin{align}
    \label{eqn:Eom1}
        0&=\partial_r^2 X+\partial_t\partial_r\Theta-\partial_t\beta, \\ 
    \label{eqn:Eom2}
        0&=\partial_t^2\Theta-\partial_r\partial_t X, \\
    \label{eqn:Eom3}
        0&=\partial_t X.
\end{align}
Varying the action also produces a boundary contribution given by
\begin{align}
    \int_{\partial\Sigma}dt\left\{-\delta X\partial_r X-\delta\Theta\partial_t X \right\}.
\end{align}
This gives us two sets of possible boundary conditions and solutions \cite{Albrychiewicz:2024tqe}. First, if both $X$ and $\Theta$ are allowed to vary freely at the endpoints, then the conditions such that the boundary term vanishes are given as 
\begin{align}
\label{eqn:NeumannCond}
    \partial_r X|_{r=0,\pi}=0, \quad \partial_t X|_{r=0,\pi}=0.
\end{align}
We refer to these as Neumann-like boundary conditions. The solutions to these equations are written as
\begin{align}
    \label{eqn:XModeExpNN}
        X(r,t)&=x_0+\sum\limits\limits_{n=1}^\infty X_n \cos(2nr), \\
    \label{eqn:YModeExpNN}    
        \Theta(r,t)&=\theta_0+\sum_{n\neq 0}\Theta_ne^{i2nr}+t\left(C+\sum\limits\limits_{n=1}^\infty2n X_n\sin(2nr)\right),
\end{align}
where $X_n$ and $\Theta_n$ are excitation modes.

The second class of boundary conditions is obtained by fixing the endpoint values of the target-space coordinates. These conditions are given as 
\begin{align}
    \partial_t X|_{r=0,\pi}=0, \quad \partial_t \Theta|_{r=0,\pi}=0,
\end{align}
with solutions 
\begin{align}
    \label{eqn:XModeExp}
        X(r,t)&=x_0+\frac{\Delta x}{\pi}r+\sum\limits_{n=1}^\infty X_n \sin(2nr), \\
    \label{eqn:YModeExp}    
        \Theta(r,t)&=\theta_0+\frac{\Delta \theta}{\pi}r+\sum\limits_{n=1}^\infty\Theta_n\sin(2nr)+t\sum\limits_{n=1}^\infty2n X_n(1-\cos(2nr)),
\end{align}
where we have defined $\Delta x=x_1-x_0$, $\Delta \theta=\theta_1-\theta_0$ with $x_0, x_1$ and $\theta_0,\theta_1$ at boundaries $r=0,\pi$ respectively. We call these Dirichlet-like boundary conditions. The canonical quantization of the Dirichlet string results in the following relations,
\begin{align}
    \left[X(r,t),P(r',t)\right]=i\delta(r-r'),\\
    \quad \left[\Theta(r,t),\Pi(r',t)\right]=i\delta(r-r').
\end{align}

The modes satisfy the following commutators,
\begin{align}
\label{eqn:ModeCan}
    \left[X_n,X_m\right]&=0,\\
    \quad \left[X_n,\Theta_m\right]&=\frac{i}{\pi}\delta_{n,m}.
\end{align}
The $X_n$ modes commute among themselves, while the $\Theta_n$ modes act as their conjugate variables. As a result, the Hamiltonian simplifies such that it only depends on the modes $X_n$ \cite{Albrychiewicz:2024tqe} 
\begin{align}
\label{eqn:HamiltonianDir}
    H=2\pi\sum\limits_{n=1}^{\infty}n^2X_n^2+\frac{\pi}{2}\left(\frac{\Delta x}{\pi}\right)^2.
\end{align}

\section{Tropical Open Strings in a Constant $B$-field}
\label{sec:Kalb_ramond_bacgrd}
We now couple the analytically continued tropical open string to a constant Kalb--Ramond background, following the standard open-string coupling to a background two-form \cite{Abouelsaood:1986gd,Seiberg:1999vs}. We choose the background field as,
\begin{align}
    B = b\, dX \wedge d\Theta,
\end{align}
where $b=B_{X\Theta}$ is constant. The worldsheet Lagrangian density for this theory is  
\begin{align}
\label{eqn:Lfull}
    \mathcal{L} = \frac{1}{2}(\partial_t \Theta)^2 - \frac{1}{2} (\partial_r X)^2 + (\beta -\partial_r\Theta)\partial_t X + b(\partial_tX\partial_r\Theta- \partial_r X \partial_t \Theta).
\end{align}
This action, too, admits the residual $\alpha$-symmetry described by \eqref{eqn:AlphaSym1}, \eqref{eqn:AlphaSym2} \cite{trsm,Albrychiewicz:2024tqe}. This follows from the fact that the action is invariant under the $\alpha$ shift of the auxiliary field $\beta$ by
\begin{equation}
\delta_\alpha \beta(t,r)=\alpha(r).
\end{equation}
The variation of the action with respect to $\alpha$ is
\begin{align}
    \delta_\alpha S = \int dt dr \alpha(r) \partial_tX,
\end{align}
Since this is a total derivative term in time, we may therefore consistently gauge-fix the $\alpha$ symmetry by imposing
\begin{align}
    \beta = 0.
\end{align}

The gauge-fixed Lagrangian density is then 
\begin{equation}
\label{eq:Lgf_tropical_B}
    \mathcal L
    =
    \frac12(\partial_t\Theta)^2
    -
    \frac12(\partial_rX)^2
    +
    (b-1)\partial_tX\,\partial_r\Theta
    -
    b\,\partial_rX\,\partial_t\Theta.
\end{equation}
These equations are independent of $b$. The Kalb--Ramond term is locally a total derivative for constant $b$ and thus its physical effect appears through the boundary conditions and the canonical structure. Varying the gauge-fixed action gives the bulk equations of motion 
\begin{align}
\label{eqn:bndyanat1}
0 &= \partial_r^2 X + \partial_t \partial_r \Theta , \\
\label{eqn:bndyanat2}
0&=\partial_t^2 \Theta - \partial_t \partial_r X .
\end{align}
The boundary contribution from the variation of the action is 
\begin{align}
\delta S\big|_{\partial_r}
= \int dt\;\Big[
\Big(-\partial_r X - b\,\partial_t\Theta\Big)\delta X
\;+\;\Big(b-1\Big)\partial_t X\,\delta\Theta
\Big]_{r=0}^{r=\pi}.
\label{eq:boundaryVariation}
\end{align}
Allowing both $\delta X$ and $\delta \Theta$ to vary freely at the endpoints results in the modified Neumann boundary conditions
\begin{align}
\partial_r X + b\,\partial_t\Theta =0,
\quad (b-1)\partial_t X=0\quad \text{at } r=0,\pi.
\label{eq:ThetaBoundaryChoice}
\end{align}
The second condition boundary condition defines two separate sectors of the theory. For generic values of $b\neq 1$, \eqref{eqn:bndyanat2} implies that

\begin{equation}
\label{endpointconstr}
    \partial_tX\big|_{\partial\Sigma}=0. 
\end{equation}
Here, the endpoints are fixed in the $X$ direction.

At $b=1$ however, the $\delta\Theta$ boundary term vanishes identically. Therefore, $\Theta$ can be left free at the boundary without imposing \eqref{endpointconstr}. The point $b=1$ defines a critical point of the theory where we must treat the generic sector $b \neq 1$ and the critical sector $b=1$ separately.  
\subsection{The Generic Sector $b\neq 1$ }
We first analyze the gauge-fixed theory for constant $b\neq 1$. The canonical momenta are defined as
\begin{align}
\label{generic_momenta_canonical}
P_X
= \frac{\partial\mathcal{L}}{\partial(\partial_t X)}
= (b-1)\,\partial_r\Theta,\\
\qquad
P_\Theta
= \frac{\partial\mathcal{L}}{\partial(\partial_t\Theta)}
= \partial_t\Theta - b\,\partial_r X.
\end{align}
Since the theory is first order in $\partial_tX$, \eqref{generic_momenta_canonical} should be understood as a primary relation between $P_X$ and $\partial_r\Theta$. When $b\neq1$, we may invert equation \eqref{generic_momenta_canonical} such that 
\begin{equation}
    \partial_r\Theta
    =
    \frac{1}{b-1}P_X .
    \label{eq:ThetaPrimePX}
\end{equation}
We now impose the standard canonical equal-time Poisson brackets on the strip $r,s\in[0,\pi]$
\begin{align}
\{X(t,r),P_X(t,s)\}=\delta(r-s)\\
\qquad
\{\Theta(t,r),P_\Theta(t,s)\}=\delta(r-s),
\end{align}
with all other brackets vanishing. Since we always work on a fixed time slice, we suppress the explicit $t$-dependence in what follows.

We proceed by reconstructing $\Theta$ from \eqref{eq:ThetaPrimePX} on $[0,\pi]$ up to its constant mode. Let $\Theta_0$ denote the zero mode of $\Theta$. Then, on the nonzero-mode space, we define the identity kernel as 
\begin{equation}
    I(r,s)
    =
    \delta(r-s)-\frac1\pi .
    \label{eq:projected_identity_generic_b}
\end{equation}
We use the standard inverse of $\partial_r$ with projector onto functions orthogonal to constants such that
\begin{align}
\Theta(r)
= \Theta_0 + \frac{1}{b-1}\int_0^\pi du\;
\Big(\theta(r-u)-\frac{r}{\pi}\Big)\,P_X(u).
\end{align}
Then the induced equal-time bracket between $X$ and $\Theta$ is
\begin{align}
\{X(r),\Theta(s)\}
&= \frac{1}{b-1}\int_0^\pi du\;
\Big(\theta(s-u)-\frac{s}{\pi}\Big)\{X(r),P_X(u)\}\nonumber\\
&= -\frac{1}{1-b}\Big(\theta(s-r)-\frac{s}{\pi}\Big).
\end{align}
Equivalently, swapping dummy variables,
\begin{align}
\{X(s),\Theta(r)\}
= -\frac{1}{1-b}\Big(\theta(r-s)-\frac{r}{\pi}\Big).
\end{align}
Under canonical quantization,
\begin{align}
\label{gencomm}
[X(s),\Theta(r)]
= i\,\{X(s),\Theta(r)\}
= -\,\frac{i}{1-b}\Big(\theta(r-s)-\frac{r}{\pi}\Big).
\end{align}
In the standard open-string case, the constant $B$-field induces a noncommutative endpoint algebra on the D-brane worldvolume \cite{Abouelsaood:1986gd,ChuHo:1998wk,Schomerus:1999ug,Seiberg:1999vs}. Here in the tropical theory, the commutator is scaled by a factor of $\frac{1}{1-b}$ and is well defined for generic $b\neq1$, but singular at $b=1$. Interestingly, it is nonlocal along the spatial directions of the open strings. However, the nonlocality of the equal-time commutator should not be interpreted as dynamical action at a distance. The $B$-field introduced only scales the canonical pairing  between $X$ and $\partial_r\Theta$, and the boundary symplectic form by a factor of $b-1$. It is thus a kinematical consequence of the first-order symplectic structure of the tropical open string.

\subsection{The Critical Sector $b=1$}

We now specialize to the critical value 
\begin{align}
    b = 1,
\end{align} where the canonical structure of the theory changes qualitatively. At first glance, one may ask whether the quantized theory exists at the limit when $b=1$ since equation \eqref{gencomm} diverges at this point. We find that this is a consequence of extra constraints and that the theory must be quantized using the Dirac--Bergmann procedure for constrained Hamiltonian systems \cite{Dirac:1950pj,Bergmann:1949zz,htbook}.

At $b=1$, the gauge-fixed Lagrangian density becomes
\begin{align}
\mathcal{L}
  = \frac{1}{2} (\partial_t \Theta)^2
    - \partial_r X\, \partial_t \Theta
    - \frac{1}{2} (\partial_r X)^2 ,
\end{align}
and the modified Neumann boundary condition at the endpoints is
\begin{align}
\Big(\partial_{r} X+\partial_t\Theta\Big)\Big|_{r = 0,\pi} &= 0.
\end{align}
The boundary variation, modified at the critical point $b=1$ in comparison with equation \eqref{eq:ThetaBoundaryChoice}, suggests that the critical point differs from the generic $b\neq 1$ theory. Away from \(b=1\), the boundary variation imposes a condition proportional to \((b-1)\partial_tX|_{\partial\Sigma}\), and hence fixes the endpoint in the \(X\)-direction. At \(b=1\), this condition vanishes identically, so the endpoint is no longer constrained to \(\partial_tX=0\). There is thus a discontinuity in the endpoint variational problem across the critical point. 

The canonical momenta are
\begin{align}
    P_X &= \frac{\partial \mathcal{L}}{\partial(\partial_t X)} = 0,\\
    P_\Theta &= \frac{\partial \mathcal{L}}{\partial(\partial_t \Theta)} = \partial_t \Theta - \partial_r X.
\end{align}
In comparison with \eqref{generic_momenta_canonical}, the momenta $P_X$  becomes non-invertible at $b=1$. Instead, it degenerates towards the  primary constraint 
\begin{align}
    \phi_1(r) = P_X(r) \approx 0,
\end{align}
where \(\approx\) denotes weak equality such that the constraint vanishes on the physical constraint surface, but should not be set to zero before evaluating Poisson brackets. Here, the endpoint decouples from fixed $X$. Thus the critical sector cannot be obtained by simply taking a $b\rightarrow 1$ limit of the Poisson brackets. Instead, it must be quantized as a constrained Hamiltonian system. 

Following the Dirac--Bergmann quantization scheme, the canonical Hamiltonian is,
\begin{align}
\label{Hcan}
H
= \int_0^\pi dr\;
\left(
\frac{1}{2}P_\Theta^2
+ P_\Theta\,\partial_r X
+ (\partial_r X)^2
\right),
\end{align}
and the total Hamiltonian is obtained by adding the primary constraint with a Lagrange multiplier,
\begin{equation}
    H_T
    =
    H
    +
    \int_0^\pi dr\,u(t,r)\phi_1(r).
\end{equation}


We impose the canonical equal-time Poisson brackets
\begin{align}
 \{X(r),P_X(s)\} &= \delta(r-s),\\
 \qquad
 \{\Theta(r),P_\Theta(s)\} &= \delta(r-s),
\end{align}
with all other elementary brackets vanishing. Since the primary constraint $\phi_1$ must be preserved under time evolution, we find that
\begin{align}
    \dot{\phi}_1(r)
    = \{\phi_1(r),H_T\}
    \approx 0,
\end{align}
which gives us the secondary constraint
\begin{align}
    \phi_2(r) = \partial_r P_\Theta(r) + 2 \partial_r^2 X(r) \approx 0.
\end{align}
No further independent constraints arise.

We now compute the constraint algebra which is encoded in the constraint matrix
\begin{align}
    C_{mn}(r,s)=\{\phi_m(r),\phi_n(s)\} .
\end{align}
The nonvanishing components come from the brackets between \(P_X\) and \(\partial_r^2X\). One thus finds 
\begin{align}
 C_{mn}(r,s)
 = \begin{pmatrix}
    0 & -\,2 \partial_r^2\delta(r-s)\\[2pt]
    2\partial_r^2\delta(r-s) & 0
   \end{pmatrix}.
\end{align}
Thus the constraints are second class, modulo the zero mode of the Neumann Laplacian. Consequently, the physical symplectic structure is not the canonical Poisson bracket on the unreduced phase space. It is the Dirac bracket obtained by inverting the constraint matrix on the subspace orthogonal to its kernel. Because $\partial_r^2$ has a Neumann zero-mode on $[0,\pi]$, an appropriate identity kernel is the projector orthogonal to constants,
\begin{align}
    I(r,s)=\delta(r-s)-\frac{1}{\pi},
\end{align}
such that we may define the Green's function $G(r,s)$ by
\begin{align}
    -\,\partial_r^2 G(r,s) \;=\; \delta(r-s)-\frac{1}{\pi},
\end{align}
with Neumann boundary conditions in the first argument,
\begin{align}
    \partial_r G(r,s)\big|_{r=0,\pi}=0.
\end{align}
The inverse constraint matrix may be chosen as
\begin{align}
    C^{12}(r,s) &= -\frac{1}{2}\,G(r,s),\\
    \qquad
    C^{21}(r,s) &= +\frac{1}{2}\,G(r,s),
\end{align}
such  that
\begin{align}
    \int_0^\pi du\, C_{ml}(r,u)C^{ln}(u,s)=\delta_m^{\ n}I(r,s).
\end{align}

The Dirac bracket between $X$ and $\Theta$ is
\begin{align}
\{X(r),\Theta(s)\}^*
&= \{X(r),\Theta(s)\}
 - \iint_0^\pi du\,dv\;
    \{X(r),\phi_1(u)\}
    C^{12}(u,v)
    \{\phi_2(v),\Theta(s)\} \nonumber\\
&= 0
 - \iint_0^\pi du\,dv\;
    \delta(r-u)\,C^{12}(u,v)\,
    \left(-\partial_v\delta(v-s)\right) \nonumber\\
&= \int_0^\pi dv\; C^{12}(r,v)\,\partial_v\delta(v-s)
= \frac{1}{2}\,\partial_s G(r,s).
\end{align}
Using the standard solution for the Green's function on $[0,\pi]$ gives
\begin{align}
\{X(r),\Theta(s)\}^*
= -\frac{1}{2}\Big(\theta(s-r)-\frac{s}{\pi}\Big).
\end{align}
Under canonical quantization, the commutator found is
\begin{align}
\label{critcomm}
    [X(r),\Theta(s)]
    = i\,\{X(r),\Theta(s)\}^*
    = -\,\frac{i}{2}\Big(\theta(s-r)-\frac{s}{\pi}\Big).
\end{align}

The commutator found here is finite and well defined, indicating that the $b=1$ theory is a separate constrained sector rather than as the naive $b\rightarrow 1$ limit of the generic theory. In this critical sector, the commutator also exhibits the same nonlocal structure as in equation \eqref{gencomm} reflected by the step function dependence. It is again a kinematical feature of the reduced equal-time phase space and should not be interpreted as dynamical propagation between the separated points. 

The appearance of the critical sector arises from the structure of the tropical theory. The degenerations of the worldsheet and target space constrains the localization equations such that transverse variations along the radial direction is coupled to motion along the leaves of the foliation. After analytic continuation, the coupling produces the term $-\partial_r\Theta \partial_tX$ in the tropical action. Yet, the background magnetic field also provides a similar coupling of the same form scaled by a factor $b$, namely
$\sim b\partial_r\Theta \partial_tX$. At the point where $b=1$, the additional coupling equipped within the tropical theory cancels that of the background magnetic field and thus produces an additional constraint on phase space since the Lagrangian no longer has an $\dot{X}$ dependence.

Though the decoupling of the endpoints at $b=1$ is different in nature from the standard Seiberg--Witten large-$B$ decoupling limit, one may still ask whether a noncommutative boundary gauge theory exists on the worldvolume of the brane. If such a boundary theory exists, the topological origin of the tropological model suggests that it may be topological rather than an ordinary dynamical noncommutative gauge theory. One may thus consider the role of A-branes, coisotropic branes, and quantization in the tropical setting \cite{Kapustin:2001ij,Gukov:2008ve,Albrychiewicz:2025hzt}. We leave this question for future work.

\section{Boundary Symplectic Structure}
\label{sec:can_sym_form}

We now rederive the noncommutative endpoint structure from the canonical
symplectic form. This provides a useful geometric interpretation of the
Poisson brackets obtained above and makes transparent where the dependence
on the Kalb--Ramond parameter \(b\) resides. We continue to work on a fixed time slice and take $b$ to be constant.

\subsection*{Mathematical Preliminaries}

Let \(\mathcal{P}\) denote the phase space of the theory, regarded formally
as an infinite-dimensional differentiable manifold, and let \(\delta\)
denote the exterior derivative on \(\mathcal{P}\). A two-form
\(\Omega\in\Omega^2(\mathcal{P})\) is symplectic if it is closed and
nondegenerate:
\begin{align}
    \delta\Omega &= 0,
    \label{eq:symplectic_closedness}\\
    \iota_V\Omega &= 0
    \quad\Longrightarrow\quad
    V=0,
    \qquad V\in T_{\zeta}\mathcal{P}.
    \label{eq:symplectic_nondegeneracy}
\end{align}
Equivalently, nondegeneracy requires the map
\(\Omega_\zeta^\flat:T_{\zeta}\mathcal{P}\rightarrow T_{\zeta}^*\mathcal{P}\), defined by
\(V\mapsto\iota_V\Omega\), to have a trivial kernel. Whenever this map is
invertible, its inverse determines the Poisson bivector and hence the
Poisson bracket.

A symplectic potential is a one-form
\(\Theta_{\mathrm{can}}\in\Omega^1(U)\), defined on an open subset
\(U\subseteq\mathcal{P}\), such that
\begin{equation}
    \Omega
    =
    \delta\Theta_{\mathrm{can}}.
    \label{eq:symplectic_form_definition}
\end{equation}
The closedness of \(\Omega\) follows from \(\delta^2=0\). The potential is
defined only up to the shift
\(\Theta_{\mathrm{can}}\rightarrow\Theta_{\mathrm{can}}+\delta F\), and a
global symplectic potential exists when \(\Omega\) is exact. For a canonical
phase space \(\mathcal{P}=T^*\mathcal{Q}\), it is the Liouville one-form,
which in local canonical coordinates \((q^i,p_i)\) is
\begin{equation}
    \Theta_{\mathrm{can}}
    =
    p_i\,\delta q^i.
\end{equation}

\subsection{Canonical Symplectic Structure of the Tropical String}

For the tropical string, the canonical variables on a fixed time slice are
\begin{equation}
    \zeta
    =
    \bigl(
        X(r),\Theta(r);
        P_X(r),P_\Theta(r)
    \bigr),
    \qquad
    r\in[0,\pi].
\end{equation}
A tangent vector \(V\in T_{\zeta}\mathcal{P}\) has components
\begin{equation}
    V
    =
    \bigl(
        V_X,V_\Theta;
        V_{P_X},V_{P_\Theta}
    \bigr).
\end{equation}
The canonical symplectic potential is defined by its action on \(V\)
\begin{equation}
    \left.
    \Theta_{\mathrm{can}}
    \right|_{\zeta}(V)
    =
    \int_0^\pi dr\,
    \left(
        P_X(r)V_X(r)
        +
        P_\Theta(r)V_\Theta(r)
    \right).
    \label{eq:canonical_symplectic_potential_intrinsic}
\end{equation}
In field-space differential notation, the canonical symplectic potential is the one-form on phase space
\begin{equation}
    \Theta_{\mathrm{can}}
    =
    \int_0^\pi dr\,
    \left(
        P_X\,\delta X
        +
        P_\Theta\,\delta\Theta
    \right).
    \label{eq:canonical_symplectic_potential}
\end{equation}
where \(\delta\) denotes the exterior derivative on field space.
Its exterior derivative symplectic two-form is
\begin{equation}
    \Omega
    =
    \delta\Theta_{\mathrm{can}}
    =
    \int_0^\pi dr\,
    \left(
        \delta P_X\wedge\delta X
        +
        \delta P_\Theta\wedge\delta\Theta
    \right).
    \label{eq:canonical_symplectic_form}
\end{equation}
The wedge product here is the antisymmetric product on field space
\begin{equation}
    \delta A\wedge\delta B
    =
    \delta A\otimes\delta B
    -
    \delta B\otimes\delta A .
\end{equation}

\subsection{Bulk--Boundary Decomposition}

For the gauge-fixed tropical string in a constant Kalb--Ramond background, the canonical momenta are
\begin{equation}
    P_X
    =
    (b-1)\partial_r\Theta
    =
    -(1-b)\partial_r\Theta,
    \qquad
    P_\Theta
    =
    \partial_t\Theta
    -
    b\,\partial_r X .
    \label{eq:generic_b_momenta_symplectic}
\end{equation}
Pulling the symplectic form back using these relations gives
\begin{equation}
    \delta P_X
    =
    -(1-b)\partial_r\delta\Theta,
    \qquad
    \delta P_\Theta
    =
    \delta(\partial_t\Theta)
    -
    b\,\partial_r\delta X .
\end{equation}
Substituting these expressions into \eqref{eq:canonical_symplectic_form}, one obtains
\begin{align}
    \Omega
    &=
    \int_0^\pi dr\,
    \left[
        -(1-b)\partial_r\delta\Theta\wedge\delta X
        +
        \delta(\partial_t\Theta)\wedge\delta\Theta
        -
        b\,\partial_r\delta X\wedge\delta\Theta
    \right].
    \label{eq:symplectic_form_before_parts}
\end{align}
Using the antisymmetry of the wedge product, the last term can be written as
\begin{equation}
    -b\,\partial_r\delta X\wedge\delta\Theta
    =
    b\,\delta\Theta\wedge\partial_r\delta X .
\end{equation}
The first term in \eqref{eq:symplectic_form_before_parts} can be integrated by parts along the spatial coordinate \(r\)
\begin{align}
    \int_0^\pi dr\,
    \left[
        -(1-b)\partial_r\delta\Theta\wedge\delta X
    \right]
    &=
    -(1-b)
    \left[
        \delta\Theta\wedge\delta X
    \right]_{0}^{\pi}
    \notag\\
    &\quad
    +(1-b)
    \int_0^\pi dr\,
    \delta\Theta\wedge\partial_r\delta X .
    \label{eq:symplectic_parts_first_term}
\end{align}
Therefore, the symplectic form becomes
\begin{align}
    \Omega
    &=
    -(1-b)
    \left[
        \delta\Theta\wedge\delta X
    \right]_0^\pi
    \notag\\
    &\quad
    +
    \int_0^\pi dr\,
    \left[
        (1-b)\delta\Theta\wedge\partial_r\delta X
        +
        b\,\delta\Theta\wedge\partial_r\delta X
        +
        \delta(\partial_t\Theta)\wedge\delta\Theta
    \right].
\end{align}
Combining the two bulk terms proportional to
$\delta\Theta\wedge\partial_r\delta X$, we arrive at the decomposition
\begin{equation}
    \Omega
    =
    \Omega_{\mathrm{bulk}}
    +
    \Omega_{\partial},
    \label{eq:symplectic_bulk_boundary_split}
\end{equation}
where
\begin{equation}
    \Omega_{\mathrm{bulk}}
    =
    \int_0^\pi dr\,
    \left[
        \delta\Theta\wedge\partial_r\delta X
        +
        \delta(\partial_t\Theta)\wedge\delta\Theta
    \right],
    \label{eq:bulk_symplectic_form}
\end{equation}
and
\begin{equation}
    \Omega_{\partial}
    =
    -(1-b)
    \left[
        \delta\Theta\wedge\delta X
    \right]_0^\pi .
    \label{eq:boundary_symplectic_form}
\end{equation}
Equivalently, the boundary contribution can be rewritten as
\begin{equation}
    \Omega_{\partial}
    =
    (1-b)
    \left[
        \delta X\wedge\delta\Theta
    \right]_0^\pi .
    \label{eq:boundary_symplectic_form_equivalent}
\end{equation}

This expression has an important interpretation. The bulk symplectic form is
independent of \(b\). All dependence on the constant Kalb--Ramond parameter
is localized on the boundary symplectic form. Thus the background
two-form does not modify the local bulk symplectic density; rather, it
changes the phase-space structure of the open-string endpoints.

\section{Conclusion}
\label{sec:con}
In this paper, we reviewed the construction of the tropical open strings and their boundary conditions. We then studied these open strings in the presence of a constant Kalb--Ramond background. Here, the familiar Seiberg--Witten mechanism \cite{Seiberg:1999vs} was used to compare the noncommutative behavior in the tropical setting. In particular, we asked how the noncommutative endpoint geometry is modified when the worldsheet and target-space geometries are degenerate and foliated. 

We find several interesting features of the tropical theory that differ from the Seiberg--Witten story. The tropical theory  already includes a coupling between the leaf direction and the transverse direction. As a result, when the Kalb--Ramond field is introduced, this term is rescaled. We ultimately find two separate sectors of the tropical theory, one for generic $b$, and one for a critical value that occurs at $b=1$. For generic values of $b\neq1$, we find that the resulting equal-time commutator between $X$ and $\Theta$ is nonlocal along the open string and is controlled by a factor of $(1-b)^{-1}$. The nonlocality, however, was found to be a kinematical consequence of the first-order symplectic structure rather than dynamical.

At the value of $b=1$, the analysis used in the generic sector no longer works. When the Kalb--Ramond field reaches a critical value such that the first-order coupling between $X$ and $\partial_r\Theta$ vanish, a primary constraint arises. Because of this extra constraint, the critical sector must be treated with the Dirac--Bergmann procedure for constrained Hamiltonian systems \cite{Dirac:1950pj,Bergmann:1949zz,htbook}. The resulting Dirac bracket is finite and well defined. 

This critical point also has an interesting interpretation of the boundary data. In the generic sector, the modified Neumann boundary conditions force the endpoints to be fixed in the $X$ direction. However, at $b = 1$, this condition vanishes and the endpoints decouple from the fixed-$X$ tropical brane. This differs from the ordinary Seiberg--Witten large-$B$ decoupling limit.

We then re-derived the endpoint noncommutativity from the canonical symplectic form. We found that the constant $B$-field doesn't introduce any new nonlocal bulk dynamics. Instead, the dependence on $b$ only scales the boundary symplectic form. In the generic case, the endpoint  Poisson structure is controlled by a factor of $(1-b)^{-1}$ and at $b=1$, the boundary symplectic form degenerates indicating the need for the Dirac--Bergmann procedure. 

Several questions remain open. It would be interesting to determine whether the critical sector admits a natural boundary gauge theory and whether such boundary gauge-theory may be topological. Such a construction would define a new foliated worldvolume theory of the branes associated to the tropical open string. It would also be interesting to clarify the relation between the noncommutative behavior of these theories and the proposed Schwinger--Keldysh wedge region \cite{neq,ssk,keq}.

\section*{Acknowledgements}
We wish to thank Andrés Franco Valiente and Emil Albrychiewicz for their help with this project. In addition, we would like to thank Petr Hořava, Ahmed Abdalla, and Sam D'Ambrosia for helpful discussions. VH is supported in part by the
Leinweber Institute for Theoretical Physics at UC Berkeley. SD is supported by the Shuimu Tsinghua Scholar Program of Tsinghua University and the Beijing Natural Science Foundation of China under Grant No.~IS25035.  

\bibliographystyle{JHEP}
\bibliography{ticfpiqm.bib}

@article{neq,
      author         = "Ho\v{r}ava, Petr and Mogni, Christopher J.",
      title          = "{Large-$N$ expansion and string theory out of equilibrium}",
 doi = "10.1103/PhysRevD.106.106013",
    journal = "Phys. Rev.",
    volume = "D106",
    number = "10",
    pages = "106013",
    year = "2022",
    eprint         = "arXiv:2008.11685",
}

@article{ssk,
    author = "Ho\v{r}ava, Petr and Mogni, Christopher J.",
    title = "{String perturbation theory on the Schwinger-Keldysh time contour}",
    eprint = "arXiv:2009.03940",
    archivePrefix = "arXiv",
    primaryClass = "hep-th",
    doi = "10.1103/PhysRevLett.125.261602",
    journal = "Phys. Rev. Lett.",
    volume = "125",
    number = "26",
    pages = "261602",
    year = "2020"
}

@article{keq,
    author = "Ho\v{r}ava, Petr and Mogni, Christopher J.",
    title = "{Keldysh rotation in the large-$N$ expansion and string theory out of equilibrium}",
    eprint = "arXiv:2010.10671",
    archivePrefix = "arXiv",
    primaryClass = "hep-th",
    doi = "10.1103/PhysRevD.106.106014",
    journal = "Phys. Rev. D",
    volume = "106",
    number = "10",
    pages = "106014",
    year = "2022"
}

@article{ewtsm,
      author         = "Witten, Edward",
      title          = "{Topological sigma models}",
      journal        = "Commun. Math. Phys.",
      volume         = "118",
      year           = "1988",
      pages          = "411",
      doi            = "10.1007/BF01466725",
      reportNumber   = "IASSNS-HEP-88/7",
      SLACcitation   = "%%CITATION = CMPHA,118,411;%%"
}

@article{trsm,
    author = "Albrychiewicz, Emil and Ellers, Kai-Isaak and Franco Valiente, Andr\'es and Ho\v{r}ava, Petr",
    title = "{Tropological sigma models}",
    eprint = "2311.00745",
    archivePrefix = "arXiv",
    primaryClass = "hep-th",
    doi = "10.1007/JHEP06(2024)135",
    journal = "JHEP",
    volume = "06",
    pages = "135",
    year = "2024"
}

@article{Albrychiewicz:2024tqe,
    author = "Albrychiewicz, Emil and Franco Valiente, Andr\'es and Hong, Vi",
    title = "{Tropical Branes}",
    eprint = "2412.12337",
    archivePrefix = "arXiv",
    primaryClass = "hep-th",
    month = "12",
    year = "2024"
}

@article{Abouelsaood:1986gd,
    author = "Abouelsaood, Ahmed and Callan, Jr., Curtis G. and Nappi, C. R. and Yost, S. A.",
    title = "{Open strings in background gauge fields}",
    reportNumber = "Print-86-1189 (PRINCETON)",
    doi = "10.1016/0550-3213(87)90164-7",
    journal = "Nucl. Phys. B",
    volume = "280",
    pages = "599--624",
    year = "1987"
}

@article{Seiberg:1999vs,
    author = "Seiberg, Nathan and Witten, Edward",
    title = "{String theory and noncommutative geometry}",
    eprint = "hep-th/9908142",
    archivePrefix = "arXiv",
    reportNumber = "IASSNS-HEP-99-74",
    doi = "10.1088/1126-6708/1999/09/032",
    journal = "JHEP",
    volume = "09",
    pages = "032",
    year = "1999"
}

@article{virohyper,
  title="{Hyperfields for tropical geometry I. Hyperfields and dequantization}",
  author="Viro, Oleg",
  eprint="arXiv:1006.3034",
  year="2010"
}

@article{viro,
  title="{On basic concepts of tropical geometry}",
  author={Viro, O. Ya.},
  journal={Proceedings of the Steklov Institute of Mathematics},
  volume={273},
  pages={252--282},
  year={2011},
  publisher={Springer}
}

@article{litvinov,
      title="{The Maslov dequantization, idempotent and tropical mathematics: A brief introduction}", 
      author={G. L. Litvinov},
      year={2005},
      eprint={arXiv:math/0507014},
      archivePrefix={arXiv},
      primaryClass={math.GM}
}

@article{Witten:1991zz,
    author = "Witten, Edward",
    editor = "Yau, Shing-Tung",
    title = "{Mirror manifolds and topological field theory}",
    eprint = "hep-th/9112056",
    archivePrefix = "arXiv",
    reportNumber = "IASSNS-HEP-91-83",
    journal = "AMS/IP Stud. Adv. Math.",
    volume = "9",
    pages = "121--160",
    year = "1998"
}

@article{Kapustin:2001ij,
    author = "Kapustin, Anton and Orlov, Dmitri",
    title = "{Remarks on A branes, mirror symmetry, and the Fukaya category}",
    eprint = "hep-th/0109098",
    archivePrefix = "arXiv",
    reportNumber = "CALT-68-2348",
    doi = "10.1016/S0393-0440(03)00026-3",
    journal = "J. Geom. Phys.",
    volume = "48",
    pages = "84",
    year = "2003"
}

@article{Gukov:2008ve,
    author = "Gukov, Sergei and Witten, Edward",
    title = "{Branes and Quantization}",
    eprint = "0809.0305",
    archivePrefix = "arXiv",
    primaryClass = "hep-th",
    doi = "10.4310/ATMP.2009.v13.n5.a5",
    journal = "Adv. Theor. Math. Phys.",
    volume = "13",
    number = "5",
    pages = "1445--1518",
    year = "2009"
}

@book{msintro,
  title={Introduction to tropical geometry, {\rm Graduate Studies in Mathematics}},
  author={Maclagan, Diane and Sturmfels, Bernd},
  volume={161},
  year={2015},
  publisher={American Mathematical Society}
}

@article{ewcoho,
      author         = "Witten, Edward",
      title          = "{Introduction to cohomological field theories}",
      booktitle      = "{Topological Methods in Quantum Field Theory Trieste,
                        Italy, June 11-15, 1990}",
      journal        = "Int. J. Mod. Phys.",
      volume         = "A6",
      year           = "1991",
      pages          = "2775-2792",
      doi            = "10.1142/S0217751X91001350",
      reportNumber   = "IASSNS-HEP-90-57",
      SLACcitation   = "%%CITATION = IMPAE,A6,2775;%%"
}

@article{Albrychiewicz:2025hzt,
    author = "Albrychiewicz, Emil and Franco Valiente, Andr\'es and Hong, Vi",
    title = "{A Tropical Look at Coisotropic Branes and Quantization}",
    eprint = "2502.19246",
    archivePrefix = "arXiv",
    primaryClass = "hep-th",
    month = "2",
    year = "2025"
}

@book{htbook,
      author         = "Henneaux, M. and Teitelboim, C.",
      title          = "{Quantization of Gauge Systems}",
      publisher      = "Princeton University Press",
      year           = "1992",
      ISBN           = "0691037698, 9780691037691",
      SLACcitation   = "%%CITATION = INSPIRE-345963;%%"
}

@article{Bergshoeff:2023rkk,
    author = "Bergshoeff, Eric and Figueroa-O'Farrill, Jos\'e and van Helden, Kevin and Rosseel, Jan and Rotko, Iisakki and ter Veldhuis, Tonnis",
    title = "{$p$-brane Galilean and Carrollian geometries and gravities}",
    eprint = "2308.12852",
    archivePrefix = "arXiv",
    primaryClass = "hep-th",
    doi = "10.1088/1751-8121/ad4c62",
    journal = "J. Phys. A",
    volume = "57",
    number = "24",
    pages = "245205",
    year = "2024"
}

@article{Kluson:2022jxh,
    author = "Kluson, J.",
    title = "{Note about D-branes in Carrollian background}",
    eprint = "2203.16824",
    archivePrefix = "arXiv",
    primaryClass = "hep-th",
    doi = "10.1007/JHEP08(2022)203",
    journal = "JHEP",
    volume = "08",
    pages = "203",
    year = "2022"
}

@article{Bergshoeff:2020xhv,
    author = "Bergshoeff, Eric and Izquierdo, Jos\'e Manuel and Romano, Luca",
    title = "{Carroll versus Galilei from a Brane Perspective}",
    eprint = "2003.03062",
    archivePrefix = "arXiv",
    primaryClass = "hep-th",
    doi = "10.1007/JHEP10(2020)066",
    journal = "JHEP",
    volume = "10",
    pages = "066",
    year = "2020"
}

@article{Dirac:1950pj,
    author = "Dirac, Paul A. M.",
    title = "{Generalized Hamiltonian dynamics}",
    doi = "10.4153/CJM-1950-012-1",
    journal = "Can. J. Math.",
    volume = "2",
    pages = "129--148",
    year = "1950"
}

@article{Bergmann:1949zz,
    author = "Bergmann, Peter G.",
    title = "{Non-Linear Field Theories}",
    doi = "10.1103/PhysRev.75.680",
    journal = "Phys. Rev.",
    volume = "75",
    pages = "680--685",
    year = "1949"
}

@article{Ardalan:1998ce,
  author = {Ardalan, F. and Arfaei, H. and Sheikh-Jabbari, M. M.},
  title = {Noncommutative Geometry from Strings and Branes},
  journal = {JHEP},
  volume = {02},
  pages = {016},
  year = {1999},
  eprint = {hep-th/9810072},
  archivePrefix = {arXiv}
}

@article{ChuHo:1998wk,
  author = {Chu, Chong-Sun and Ho, Pei-Ming},
  title = {Noncommutative Open String and D-brane},
  journal = {Nucl. Phys. B},
  volume = {550},
  pages = {151--168},
  year = {1999},
  eprint = {hep-th/9812219},
  archivePrefix = {arXiv}
}

@article{Schomerus:1999ug,
  author = {Schomerus, Volker},
  title = {D-branes and Deformation Quantization},
  journal = {JHEP},
  volume = {06},
  pages = {030},
  year = {1999},
  eprint = {hep-th/9903205},
  archivePrefix = {arXiv}
}

@article{Buzaglo:2025bcca,
  author  = {Buzaglo, Lucas and He, Xiao and Pham, Tuan Anh and Tan, Haijun and Vishwa, Girish S. and Zhao, Kaiming},
  title   = {On the boundary Carrollian conformal algebra},
  journal = {Letters in Mathematical Physics},
  volume  = {116},
  number  = {86},
  year    = {2026},
  doi     = {10.1007/s11005-026-02118-z},
  eprint  = {2508.21603},
  archivePrefix = {arXiv},
  primaryClass = {math.RT}
}

@article{Duary:2025hdb,
    author = "Duary, Sarthak and Maji, Sourav",
    title = "{From closed to open strings: the tensionless route in Kalb-Ramond background and noncommutativity}",
    eprint = "2511.20917",
    archivePrefix = "arXiv",
    primaryClass = "hep-th",
    month = "11",
    year = "2025"
}

\end{document}